\pdfoutput=1

\documentclass[11pt]{article}
\usepackage{booktabs}
\usepackage{longtable}
\usepackage{amsmath}
\usepackage{booktabs}
\usepackage[final]{acl}
\usepackage{hyperref}
\usepackage{times}
\usepackage{latexsym}
\usepackage[utf8]{inputenc}
\usepackage[T2A]{fontenc}
\usepackage[russian,english]{babel}
\usepackage{multirow}
\usepackage[most]{tcolorbox}
\usepackage{xcolor} 

\usepackage[T1]{fontenc}

\usepackage[utf8]{inputenc}

\usepackage{microtype}

\usepackage{inconsolata}

\usepackage{graphicx}

\title{Rethinking Search: A Study of University Students’ Perspectives on Using LLMs and Traditional Search Engines in Academic Problem Solving}

\author{
 \textbf{Md. Faiyaz Abdullah Sayeedi\textsuperscript{1}},
 \textbf{Md. Sadman Haque\textsuperscript{1}},
 \textbf{Zobaer Ibn Razzaque\textsuperscript{1}},\\
 \textbf{Robiul Awoul Robin\textsuperscript{1}},
 \textbf{Sabila Nawshin\textsuperscript{2}}
\\
\\
 \textsuperscript{1}United International University, Bangladesh \qquad
 \textsuperscript{2}Indiana University Bloomington, USA
\\
 \small{
   \textbf{Emails:} \texttt{\{msayeedi212049,mhaque221592,zrazzaque221135,rrobin221564\}@bscse.uiu.ac.bd}; \texttt{snawshin@iu.edu}
 }
}

\begin{document}
\maketitle
\begin{abstract}
With the increasing integration of Artificial Intelligence (AI) in academic problem solving, university students frequently alternate between traditional search engines like Google and large language models (LLMs) for information retrieval. This study explores students’ perceptions of both tools, emphasizing usability, efficiency, and their integration into academic workflows. Employing a mixed-methods approach, we surveyed 109 students from diverse disciplines and conducted in-depth interviews with 12 participants. Quantitative analyses, including ANOVA and chi-square tests, were used to assess differences in efficiency, satisfaction, and tool preference. Qualitative insights revealed that students commonly switch between GPT and Google: using Google for credible, multi-source information and GPT for summarization, explanation, and drafting. While neither tool proved sufficient on its own, there was a strong demand for a hybrid solution. In response, we developed a prototype, a chatbot embedded within the search interface, that combines GPT’s conversational capabilities with Google’s reliability to enhance academic research and reduce cognitive load.
\end{abstract}

\section{Introduction}

The rapid advancement of artificial intelligence has significantly reshaped the ways in which university students seek academic information and engage in research activities \citep{pirzado2024navigating}. Traditionally, search engines like Google have served as the dominant tool for retrieving scholarly content due to their accessibility, breadth of indexed materials, and access to verified sources. However, the emergence of LLMs, such as OpenAI's ChatGPT, has introduced a new paradigm—offering students direct, conversational responses and contextualized summaries that can streamline information consumption \citep{alberth2023use}.

This evolution in digital research tools raises important questions about how students perceive and utilize these systems, particularly in academic settings where accuracy, credibility, and efficiency are critical. Prior research suggests that while LLMs facilitate rapid content summarization and task-specific assistance, their reliability varies depending on context and task complexity \citep{divekar2024choosing, xu2023chatgpt}. Conversely, search engines provide access to a wide range of authoritative sources but often require users to sift through multiple links and evaluate conflicting information independently. Several studies have documented this complementary behavior—students tend to use LLMs for explanation and drafting while relying on search engines for fact-checking and source validation \citep{caramancion2024large, spatharioti2023comparing}.

Despite their respective strengths, both tools also have well-documented limitations. LLMs can generate confident yet incorrect outputs, potentially misleading users \citep{xu2023chatgpt}, while traditional search engines can lead to information overload and inefficiency in time-sensitive academic contexts. As a result, students are increasingly adopting a hybrid approach—strategically switching between LLMs and search engines to balance speed with credibility \citep{sakirin2023user, kapoor2024ai}. However, this constant toggling between tools introduces cognitive overhead and fragmented workflows, especially when performing complex academic tasks.

To investigate these dynamics systematically, this study addresses the following research questions:

\begin{itemize}
    \item \textbf{RQ1:} How do university students perceive the usability, efficiency, and satisfaction of LLMs compared to traditional search engines in academic problem-solving?
    \item \textbf{RQ2:} What patterns of tool usage emerge when students perform academic tasks with either or both tools?
    \item \textbf{RQ3:} What are students’ preferences and expectations for an integrated solution that combines the strengths of both systems?
\end{itemize}

\section{Literature Review}

LLMs have significantly reshaped how individuals learn, make decisions, and retrieve information. While traditional search engines like Google have long been the primary tool for academic information seeking, recent research increasingly explores how LLMs compare in terms of usability, task performance, and user trust. \citet{divekar2024choosing} examined how university students use LLMs like ChatGPT alongside traditional search engines for learning new topics. Their findings indicate that while LLMs support rapid summarization and ease of understanding, their effectiveness varies depending on the complexity and nature of the task. In a similar vein, \citet{kumar2024understanding} analyzed how students use LLMs to generate SQL queries. They observed that LLM assistance improved query formulation and contributed positively to the learning experience.

Several studies have also investigated task completion performance. \citet{spatharioti2023comparing} conducted a randomized experiment and found that LLM users completed decision-making tasks more quickly and with fewer queries. However, the authors warned of a major drawback: users often overtrust LLM outputs, especially when incorrect answers are presented confidently. They suggested the inclusion of confidence indicators to mitigate this issue. \citet{xu2023chatgpt} echoed this concern, emphasizing the need for rigorous fact-checking when relying on LLM responses.

In terms of task preference, \citet{caramancion2024large} evaluated 20 types of information-seeking scenarios and concluded that users favored traditional search engines for fact-based queries, while preferring LLMs for creative or complex tasks. Supporting this, \citet{sakirin2023user} found that nearly 70\% of participants preferred ChatGPT-style conversational interfaces due to their personalization, perceived efficiency, and convenience. Extending these findings, \citet{wazzan2024comparing} studied image geolocation tasks and observed that tool selection often influenced user strategy: LLMs were used more intuitively, while traditional tools required structured navigation.

The issue of credibility remains central. \citet{kapoor2024ai} argued that despite the convenience and rapidity of AI tools, traditional search methods remain more reliable for academic research. In contrast, LLMs often lack source transparency, which can be problematic in scholarly settings. To address this trade-off, researchers have proposed hybrid models. \citet{bal2009comparative} explored metasearch engines that aggregate content from various sources to improve accuracy, and \citet{caramancion2024large} advocated for systems that combine the contextual depth of LLMs with the source validation strengths of search engines.

However, existing studies have primarily evaluated LLMs and traditional search engines in isolation or through task-specific comparisons, without fully exploring how students naturally combine both tools in academic workflows \citep{xu2023chatgpt}. There is a lack of empirical research that integrates both performance metrics and user perspectives to understand this hybrid usage behavior \citep{bansal2023optimizing}. While tools like Perplexity AI \citep{perplexityai2024} attempt to bridge this gap by combining AI-generated responses with source links, and Google has introduced AI summaries through its Search Generative Experience (SGE), these systems remain largely static, lacking personalization, real-time adaptation, and task-specific reasoning. This study addresses these limitations through a mixed-methods approach and the design of a user-informed, context-aware prototype.

\section{Methodology}

\begin{figure*}[ht]
  \centering
  \includegraphics[width=\textwidth]{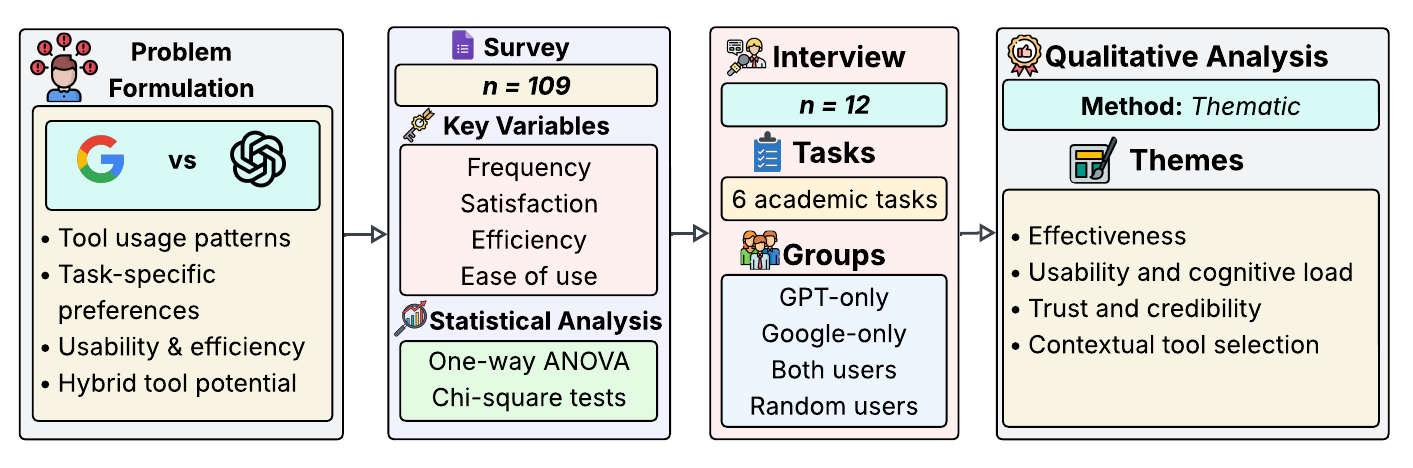}
  \caption{
    Overview of the study methodology. A mixed-methods approach was employed in this study. (1) The survey phase $(n = 109)$ captured quantitative data and analyzed using different statistical tests. (2) The qualitative phase included in-person interviews $(n = 12)$, where participants completed six academic tasks and were grouped based on tool usage. Thematic analysis of open-ended responses and interview transcripts led to four themes.
  }
  \label{fig:methodology}
\end{figure*}

To explore university students’ preferences and usage behaviors regarding LLMs and traditional search engines for academic tasks, we employed a mixed-methods approach that combined quantitative and qualitative data collection and analysis, shown in Figure \ref{fig:methodology}. This design allowed us to examine both broad patterns and deeper user experiences in a complementary manner.

We first conducted an online survey that collected responses from 109 university students across a range of academic disciplines. While the survey primarily targeted students in technology-related fields such as Computer Science and Engineering (CSE), Electrical and Electronics Engineering (EEE), and Data Science, it was also distributed to students from other areas, including Business Administration and Medicine, to ensure diversity. The questionnaire included both closed- and open-ended questions designed to assess tool usage frequency, satisfaction, efficiency, and perceived ease of use when using GPT-based LLMs and traditional search engines like Google. Descriptive statistics were used to summarize the data, and inferential statistical tests, one-way ANOVA and chi-square tests, were employed to evaluate differences in user perceptions and the influence of demographic variables on tool preference.

To enrich and validate the survey findings, we conducted in-depth in-person interviews with 12 students from CSE, EEE, and BBA backgrounds, primarily recruited from United International University (UIU), Dhaka, Bangladesh. Each participant was asked to complete six academic tasks: summarizing a research article, solving a coding problem (only for CSE students), addressing a circuit-related issue (only for EEE students), analyzing business data (only for BBA students), drafting a formal academic email, and comparing two popular academic concepts. Participants were divided into four groups based on their tool usage behavior: GPT-only users, Google-only users, balanced users who alternated between both tools, and random-choice users who freely switched between GPT and Google depending on preference.

The qualitative analysis synthesized insights from both the open-ended survey responses and interview transcripts. Thematic coding focused on perceived effectiveness, usability, trustworthiness, and the contextual factors that influenced tool selection. This analysis provided a comprehensive understanding of how students navigate the strengths and limitations of both tools and how their choices are shaped by the nature of the academic task, familiarity with the subject, and perceived cognitive effort.

\section{Demographics}

\subsection{Survey}

The survey included 109 participants from a range of academic disciplines and demographic backgrounds. While the majority of participants came from CSE, the sample also included students from EEE, BBA, Data Science, Mathematics, Biochemistry and Biotechnology, and Medicine. Table~\ref{tab:distribution} summarizes the distribution by department, CGPA, gender, and age group.

\begin{table}[h]
  \centering
  \small
  \begin{tabular}{p{5cm}p{0.5cm}}
    \hline
    \multicolumn{2}{c}{\textbf{Department}} \\ \hline
    Computer Science and Engineering (CSE) & 72 \\ \hline
    Electrical and Electronics Engineering (EEE) & 13 \\ \hline
    Bachelor of Business Administration (BBA) & 10 \\ \hline
    Data Science & 8 \\ \hline
    Mathematics & 4 \\ \hline
    Biochemistry and Biotechnology & 1 \\ \hline
    Bachelor of Medicine, Bachelor of Surgery (MBBS) & 1 \\ \hline
    \multicolumn{2}{c}{\textbf{CGPA Range}} \\ \hline
    3.81 -- 4.00 & 23 \\ \hline
    3.51 -- 3.80 & 30 \\ \hline
    3.01 -- 3.50 & 36 \\ \hline
    2.50 -- 3.00 & 16 \\ \hline
    Below 2.50 & 4 \\ \hline
    \multicolumn{2}{c}{\textbf{Gender}} \\ \hline
    Male & 81 \\ \hline
    Female & 28 \\ \hline
    \multicolumn{2}{c}{\textbf{Age Range}} \\ \hline
    18 -- 20 years & 5 \\ \hline
    21 -- 25 years & 95 \\ \hline
    26 -- 30 years & 9 \\ \hline
  \end{tabular}
  \caption{Distribution of participants by department, CGPA range, gender, and age group.}
  \label{tab:distribution}
\end{table}

The survey instrument included Likert-scale questions assessing perceptions of traditional search engines (e.g., Google) and LLM-based tools (e.g., ChatGPT). Participants responded on a 5-point scale: \textit{Never}, \textit{Rarely}, \textit{Occasionally}, \textit{Frequently}, and \textit{Always}. Each set of questions was repeated for both tool categories, covering four core dimensions:

\begin{itemize}
  \item \textit{How often do you use the following tools for academic tasks?}
  \item \textit{How satisfied are you with the accuracy of information provided by the following tools?}
  \item \textit{How efficient are these tools in helping you complete academic tasks?}
  \item \textit{How easy are these tools to use for academic purposes?}
\end{itemize}

Participants rated these items separately for both traditional search engines and LLM-based tools. At the end of the survey, participants were also asked to indicate their overall preference.

This combination of parallel metrics and comparative judgment allowed for consistent statistical comparisons across tools, while the final preference item offered insight into holistic user inclinations.

\subsection{In-person Interview}

To complement the survey findings and provide deeper insights into tool-related behaviors, we conducted in-person interviews with 12 students from varied academic backgrounds, primarily from CSE, EEE, and BBA programs. The interview protocol included a structured sequence of six academic tasks, designed to simulate common university-level activities: (1) summarizing a research article, (2) solving a coding problem (for CSE students), (3) answering a circuit-related question (for EEE students), (4) analyzing business data (for BBA students), (5) drafting a formal email, and (6) comparing two popular academic concepts. These tasks were selected based on consultations with domain instructors and a review of typical coursework assignments, ensuring contextual relevance and varying cognitive demands. The goal was to observe how tool choice affected task strategy, accuracy, and efficiency across both discipline-specific and general academic activities.

To assess performance, we developed a task-specific rubric in consultation with faculty members in relevant fields. For example, the coding task was evaluated based on correctness and code clarity; the summarization task was scored on coherence, coverage, and conciseness; and the comparison task was assessed for clarity of distinctions and logical reasoning. Each task was scored independently by two evaluators to ensure inter-rater reliability.

Participants were divided into four groups based on their tool usage patterns during task completion: (1) GPT-only users, (2) Google-only users, (3) tool-balancing users (who used both tools sequentially), and (4) random-choice users (who selected tools freely for each task). This grouping was used to compare differences in accuracy and completion time across task types and to explore how tool-switching behavior aligned with user preferences and task complexity. The interviews also included open-ended reflections on tool usability, trust, and perceived strengths or limitations. These qualitative responses were thematically analyzed to supplement quantitative trends and inform design recommendations.

\section{Quantitative Analysis}

\subsection{Survey Results}

Closed-ended survey responses were converted into numerical values for analysis, using a 5-point Likert scale coded as follows: \textit{Never (0)}, \textit{Rarely (1)}, \textit{Occasionally (2)}, \textit{Frequently (3)}, and \textit{Always (4)}. This enabled calculation of means, medians, modes, and standard deviations across four core dimensions: usage frequency, satisfaction, efficiency, and ease of use, for both traditional search engines and LLM-based tools.

The analysis revealed that LLM-based tools were used more frequently and received more favorable ratings across all metrics. The mean usage frequency for traditional search engines was 2.33, with a median of 2.0 and a mode of 2, suggesting occasional use among participants. In contrast, LLM-based tools had a higher mean frequency of 2.79, with a median of 3.0 and a mode of 3, indicating more frequent use. Satisfaction with traditional search engines yielded a mean of 1.99, a median of 2.0, and a mode of 2, reflecting a generally neutral to slightly unsatisfied experience. LLM-based tools, on the other hand, had a higher satisfaction mean of 2.47, with a median and mode of 3.0, indicating moderate satisfaction. Standard deviations for both tools were around 0.9, suggesting consistency in responses.

\begin{figure}[h]
  \centering
  \includegraphics[width=\columnwidth]{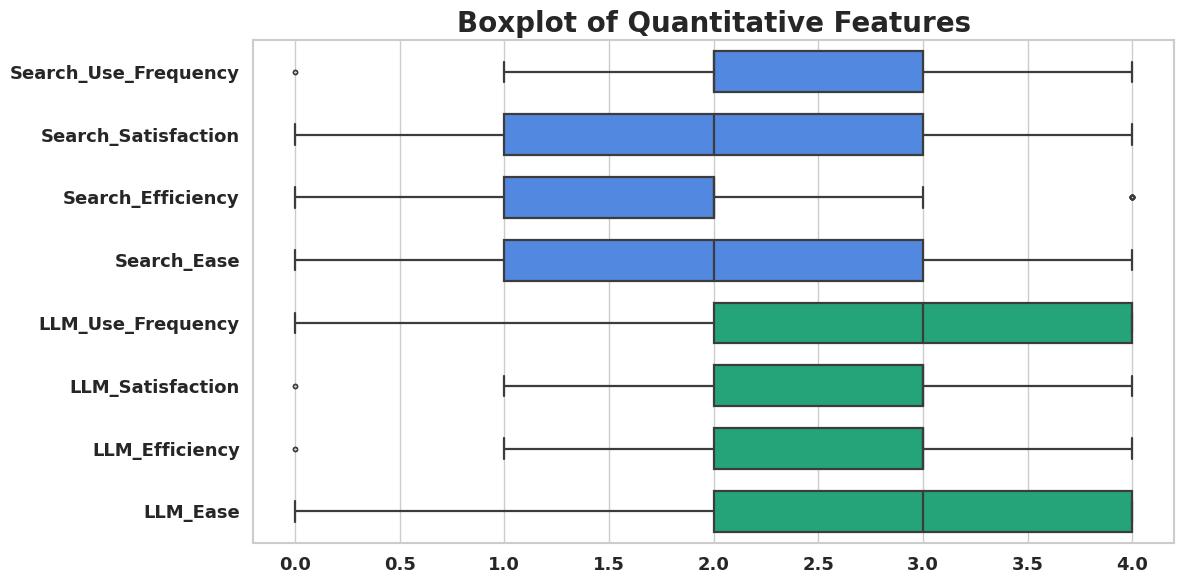}
  \caption{Boxplot of Quantitative Features: This figure presents a comparative analysis of key usability factors between traditional search engines and LLM-based tools. The top four features—\textit{Search\_Use\_Frequency}, \textit{Search\_Satisfaction}, \textit{Search\_Efficiency}, and \textit{Search\_Ease}—represent user responses related to traditional search engines. The bottom four—\textit{LLM\_Use\_Frequency}, \textit{LLM\_Satisfaction}, \textit{LLM\_Efficiency}, and \textit{LLM\_Ease}—correspond to user experiences with large language models.}
  \label{fig:boxplot}
\end{figure}

Efficiency ratings followed a similar trend. Traditional search engines received a mean score of 2.06 (median = 2.0, mode = 2), whereas LLM-based tools were perceived as more efficient, with a mean of 2.65, median of 3.0, and mode of 2. The variability in responses was moderate for both tools, with standard deviations of 0.94 and 1.00, respectively. In terms of ease of use, traditional search engines had a mean score of 2.10, a median of 2.0, and a mode of 2. LLM-based tools again outperformed, with a mean of 2.74, a median of 3.0, and a mode of 3. The standard deviation for LLM ease of use (1.12) was slightly higher, reflecting greater variability in responses. These comparisons are visually presented in Figure~\ref{fig:boxplot}.

To assess whether the observed differences in user perceptions between traditional search engines and LLM-based tools were statistically significant, we conducted a series of one-way repeated measures ANOVA tests across four dimensions: usage frequency, satisfaction, efficiency, and ease of use. This within-subjects design was appropriate as each participant rated both tools, allowing direct comparison of matched responses. The results revealed significant differences in all cases: usage frequency, $F(1, 108) = 14.82$, $p < 0.001$; satisfaction, $F(1, 108) = 18.95$, $p < 0.001$; efficiency, $F(1, 108) = 21.37$, $p < 0.001$; and ease of use, $F(1, 108) = 17.04$, $p < 0.001$. These findings indicate that participants consistently rated LLM-based tools higher than traditional search engines across all usability dimensions. All F-values were positive, as expected in ANOVA, and each test was independently conducted per variable. Assumptions of normality and sphericity were evaluated and satisfied, supporting the reliability of the results. Overall, the statistical evidence confirms that the differences in ratings are not due to chance but reflect a significant and consistent user preference for LLM-based tools in academic contexts.

To explore whether tool preference was influenced by participant background, a chi-square test was performed to examine associations between tool preference (LLM, search engine, or both) and demographic variables such as age group, gender, and academic department. The chi-squared statistic was $\chi^2$(6, N=109) = 2.012, with a p-value of 0.570. Since the p-value exceeds the 0.05 threshold, we fail to reject the null hypothesis. This indicates that tool preference is not significantly associated with any of the demographic factors analyzed.

In summary, the survey results demonstrate that participants generally prefer LLM-based tools over traditional search engines across all major dimensions of usability. While individual backgrounds, such as department or gender, did not significantly influence this preference, the performance gap between the two tools was consistently supported by both descriptive and inferential statistical analysis.

\subsection{In-person Interview}

To analyze the data collected from the in-person interviews, we examined both quantitative and qualitative aspects of participant performance while completing a series of structured academic tasks. A total of 12 students participated in this phase of the study. They were assigned six academic tasks representative of common university-level activities. These tasks were selected to reflect a range of cognitive demands, from analytical reasoning to written communication. Participants were grouped based on their tool usage strategy: GPT-only users, Google-only users, tool-balancing users (who used both tools sequentially), and random-choice users (who selected tools freely at each step). Table \ref{tab:interview_summary} summarizes the key quantitative findings from your in-person interview.

\begin{table}[h]
\centering
\small
\begin{tabular}{p{2.5cm}cc}
\hline
\textbf{Tool Usage Group} & \textbf{Accuracy (\%)} & \textbf{Time (min)} \\ \hline
GPT-only            & 83                             & 19                                     \\ \hline
Google-only         & 78                             & 24                                     \\ \hline
Tool-balancing      & 90                             & 30                                     \\ \hline
Random-choice       & 82--88                         & 22--29                                 \\ \hline
\end{tabular}
\caption{Summary of task performance by tool usage group. Accuracy (\%) refers to the average task score based on a predefined rubric. Time (min) indicates the average completion time across six academic tasks.}
\label{tab:interview_summary}
\end{table}

\subsubsection{Accuracy Analysis}

Each participant’s response was manually evaluated using a predefined scoring rubric tailored to each task type. For instance, the coding task was assessed based on functional correctness and code readability, while the summarization task was rated on coverage, conciseness, and coherence. The rubric ensured consistency and objectivity across evaluations. Participants who relied exclusively on GPT achieved an average accuracy of 83\%, suggesting that LLMs were effective in generating structured responses, particularly for summarization and drafting. In contrast, Google-only users attained an average accuracy of 78\%, likely due to the additional effort required to navigate, synthesize, and rephrase content from multiple sources. Participants who employed both tools in a balanced, complementary fashion demonstrated the highest performance, averaging 90\% accuracy. Their use of GPT for synthesis and Google for verification allowed for improved reliability and content quality. Among the eight random-choice users, who selected tools freely based on task needs, accuracy ranged from 82\% to 88\%, depending on the complexity of the task and the appropriateness of tool selection.

\subsubsection{Completion Time Analysis}

We also recorded the time taken by each participant to complete the assigned tasks. The completion times ranged from a minimum of 13 minutes to a maximum of 42 minutes across all participants. On average, GPT-only users completed tasks the fastest, requiring approximately 19 minutes. This efficiency can be attributed to the conversational nature of LLMs, which reduces the need to browse multiple webpages. Google-only users required more time, around 24 minutes on average, due to the iterative process of selecting, reading, and extracting relevant content from diverse sources. Participants who used both tools took the longest, with an average completion time of 30 minutes. However, this group also achieved the highest accuracy, suggesting a trade-off between speed and performance. The random-choice group showed the most variability in completion time, ranging from 22 to 29 minutes. Their timing appeared to be influenced by both task complexity and personal familiarity with the chosen tools. In general, the results indicate that while GPT-based tools provide speed and ease of access, combining them with traditional search engines can lead to improved accuracy, albeit at the cost of increased task duration.

\section{Qualitative Analysis}

The qualitative analysis draws on open-ended survey responses and in-person interview transcripts to explore participants’ perceptions, experiences, and decision-making strategies when using GPT and Google for academic tasks. We employed a thematic analysis approach to identify recurring patterns and categories in the qualitative data. Initial coding was conducted independently by two researchers who reviewed all textual responses line-by-line. Codes were then grouped into broader themes through iterative comparison and refinement until consensus was reached.

Four major themes emerged from the data: (1) task suitability and tool preference, (2) perceptions of reliability and accuracy, (3) workflow efficiency and cognitive load, and (4) usability and interaction experience.

\paragraph{Task Suitability and Tool Preference.} Participants frequently distinguished between tools based on the academic task. GPT was consistently described as effective for quick answers, summarization, and writing support. One participant noted, \textit{"I use ChatGPT whenever I need to summarize something quickly or generate a draft; it saves a lot of time."} In contrast, Google was preferred for tasks requiring deeper exploration and source triangulation. For example, a BBA student shared \textit{"Google helps me see what different sources are saying, especially when I need to analyze business trends from multiple angles."}

\paragraph{Perceptions of Reliability and Accuracy.} Trust emerged as a key factor in tool selection. While GPT was appreciated for its fluency and coherence, several participants expressed concerns about outdated or generalized responses. One remarked, \textit{"Sometimes GPT gives an answer that sounds right but isn’t actually correct, so I double-check with Google."} Google was consistently rated as more trustworthy for fact-checking and citing sources, though some respondents reported difficulty in assessing source quality or encountering contradictory information.

\paragraph{Workflow Efficiency and Cognitive Load.} Many participants described GPT as a way to streamline academic tasks, particularly under time pressure. For instance, a CSE student commented, \textit{"Instead of going through five different websites, I just ask GPT and get a concise answer."} However, this benefit was counterbalanced by reports of multitool use. Students who used both GPT and Google acknowledged that switching between them increased task duration but ultimately improved their understanding and output quality. This dual strategy was especially common for tasks involving coding, data analysis, or structured writing.

\paragraph{Usability and Interaction Experience.} GPT was often framed as a conversational assistant or “personal tutor” that guided the student through a problem interactively. In contrast, Google was seen as more traditional but stable. As one student described, \textit{"ChatGPT feels like someone is explaining things to me, but with Google I have to do all the work to find and compare stuff."} Interface familiarity and preferred mode of information delivery influenced tool preference, particularly for students less comfortable with long-form search or unfamiliar domains.

Overall, students perceived GPT and Google not as competing tools but as complementary components of their academic workflow. GPT was favored for its speed, language generation, and summarization abilities, while Google remained essential for verifying facts and consulting credible sources. The choice of tool depended largely on the type of task, the user’s prior knowledge, and their need for either convenience or verification. These findings highlight the nuanced, context-dependent strategies students adopt when navigating digital information tools.

\section{Discussion}

The findings from both the survey and in-person interviews reveal a nuanced interplay between LLMs and traditional search engines in academic information-seeking behavior. GPT-based systems were consistently valued for their ability to provide structured, coherent, and contextually relevant responses. Their strengths were particularly evident in tasks requiring rapid summarization, coding support, or written content generation, where participants appreciated the speed and reduced cognitive effort offered by conversational interfaces. However, while LLMs excelled in usability and perceived efficiency, their limitations, such as occasionally outdated or overly generalized content, prompted students to cross-reference with more authoritative sources.

In contrast, traditional search engines like Google remained the preferred tool for in-depth research, source validation, and academic rigor. Students highlighted Google’s extensive access to peer-reviewed literature, academic websites, and multiple viewpoints as vital for tasks requiring critical evaluation or citation. Nonetheless, participants also reported experiencing information overload and inefficiencies due to the need to manually sift through links, assess credibility, and synthesize fragmented content. These trade-offs suggest that tool preference is not static but shaped by the academic task’s complexity, time constraints, and the student’s familiarity with the subject matter.

\begin{figure}[h]
  \centering
  \includegraphics[width=0.9\columnwidth]{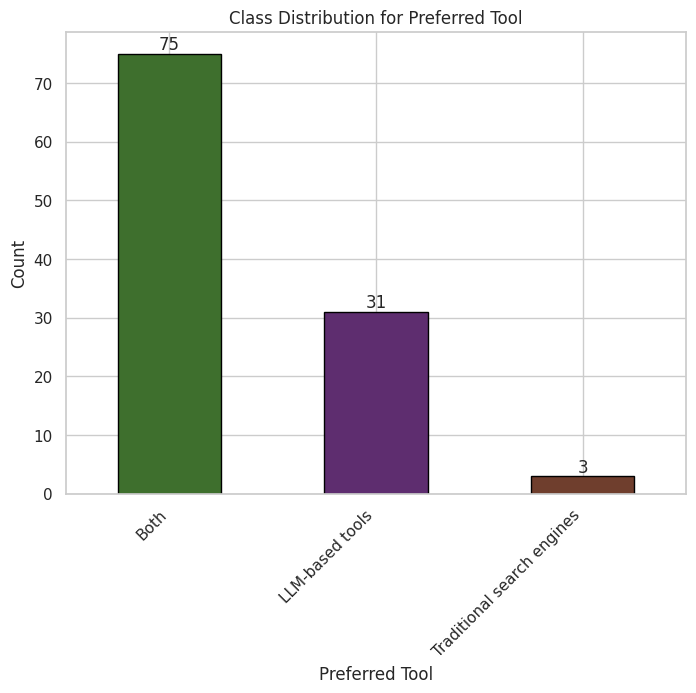}
  \caption{Class Distribution of the Preferred Tool among the Students}
  \label{fig:preferred_tool}
\end{figure}

A recurring theme across both quantitative and qualitative data was the strong interest in a hybrid model that seamlessly integrates the complementary strengths of GPT and Google. As illustrated in Figure~\ref{fig:preferred_tool}, a significant portion of participants expressed a desire for an academic support tool that combines GPT’s conversational and summarization capabilities with Google’s multi-source, real-time information retrieval. Such a system would enable users to receive concise responses with embedded citations and links to original sources, streamlining the verification process without sacrificing depth or credibility. Participants viewed this hybrid approach as a way to reduce cognitive load, eliminate repetitive tool-switching, and enhance learning outcomes through more fluid academic workflows.

While LLMs and traditional search engines serve distinct purposes, students view them as complementary rather than competing tools. The integration of their respective advantages, LLMs for generation and synthesis, and search engines for depth and verification, represents a promising direction for the future of academic information retrieval. These findings underscore the importance of designing intelligent, context-aware tools that adapt to students’ diverse needs while upholding standards of reliability and academic integrity.

These findings also raise a key question: \textit{\textbf{What happens when students use GPT and Google together: Does it help or hurt?}} The evidence from our study suggests a compelling answer. Participants who used both tools, using GPT to quickly summarize and clarify complex topics, and Google to verify facts and consult authoritative sources, consistently outperformed those who relied on either tool alone. These hybrid users achieved the highest task accuracy (90\%), demonstrating that the strategic integration of LLMs and traditional search engines not only complements their respective strengths but also minimizes their individual weaknesses. Although this dual-tool approach required more time, students perceived it as a worthwhile trade-off for greater confidence, deeper understanding, and higher-quality outcomes. This finding underscores the potential of thoughtfully designed hybrid systems to support academic workflows, reducing cognitive load while maintaining rigor and trustworthiness in the learning process.

\section{Proposed Prototype}

Drawing on user insights from both survey responses and interviews, we propose a conceptual prototype that integrates GPT-based assistance directly into the traditional search engine interface.\footnote{\href{https://www.figma.com/proto/l2bXNzMIHNqiDiicnJgL7g/HCI-Prototype?node-id=1-451&starting-point-node-id=1\%3A451}{Figma Design}} The goal is to address the cognitive and logistical burden of switching between tools by creating a unified platform that combines the conversational utility of LLMs with the source-rich infrastructure of search engines. The prototype is designed as an embedded chatbot, positioned unobtrusively in the corner of the search interface, allowing users to engage in interactive, context-aware dialogue without disrupting their familiar browsing workflow.

Unlike standalone LLM interfaces, the proposed assistant does not replace standard search results. Instead, it complements them by offering real-time summaries, follow-up clarifications, and cross-source syntheses derived from retrieved documents. For example, when a user performs a Google search, the assistant can instantly summarize key points from the top-ranked results, offer bullet-point comparisons across sources, or simplify complex academic texts. Users can also ask follow-up questions to refine or extend the information, eliminating the need to manually revisit and interpret multiple pages.

The prototype’s novelty lies in its hybrid architecture that allows toggling between raw search content and AI-enhanced interpretation. Crucially, each AI-generated insight is accompanied by links to the original source, promoting transparency and reducing the risk of hallucinated or unverifiable responses. This feature directly addresses the concern, voiced by multiple participants, regarding LLM trustworthiness in academic work.

Although conceptual in its current form, the prototype was informed by empirical findings from this study and inspired by real-world user preferences. Its contribution lies in reimagining academic information retrieval as an interactive and adaptive process. Over time, the assistant could learn user preferences, discipline-specific language, and search habits to deliver more relevant and personalized guidance. By embedding this intelligent layer into the search experience, the prototype aims to reduce cognitive load, increase search efficiency, and promote evidence-based academic practices, ultimately bridging the current gap between generation and verification in digital research tools.

\section{Conclusion}
Our study demonstrates that university students adopt a complementary approach to academic information retrieval, using LLMs for quick explanations and drafting, and traditional search engines for verification and accessing credible sources. Our mixed-methods findings underscore the task-dependent nature of tool preference and reveal strong interest in a hybrid model. While the proposed prototype remains conceptual and the interview sample was limited, the results offer practical insights for designing AI-assisted academic tools. Future work will focus on expanding participant diversity, validating qualitative themes, and implementing a functional prototype to assess real-world usability and impact.

\section*{Acknowledgments}
We sincerely thank everyone who participated in the survey and interviews for their time and experience sharing.

\section*{Limitations}
This study provides important insights into students' use of LLMs and traditional search engines for academic tasks; however, several limitations should be acknowledged. The in-person interview sample was relatively small $(n = 12)$ and primarily drawn from technology-related disciplines, which may limit the generalizability of the findings. There is also potential sampling bias, as the majority of participants were Computer Science and Engineering (CSE) majors from a single university context, which may not reflect the experiences or preferences of students from other academic backgrounds or institutions. While the qualitative analysis surfaced meaningful themes, it lacked inter-coder reliability checks, and the survey relied on self-reported data that may be affected by recall or social desirability bias. Moreover, the proposed prototype remains at a conceptual stage and has not yet been implemented or tested in real academic environments, leaving its practical impact unverified. Lastly, the study did not account for factors such as digital literacy, prior experience with AI tools, or task complexity, all of which could influence tool preferences and performance.

\section*{Ethical Considerations}

All procedures involving human participants in this study were conducted in accordance with ethical research standards. Participation in both the survey and in-person interviews was voluntary, and informed consent was obtained from all participants prior to data collection. Respondents were assured of anonymity and confidentiality, and no personally identifiable information was collected or stored. The data were used solely for research purposes and analyzed in aggregate to protect individual identities. As the study did not involve vulnerable populations, clinical interventions, or sensitive topics, risk to participants was minimal. The conceptual prototype proposed in this study does not process real user data and poses no immediate privacy concerns. Future implementation of the prototype will incorporate robust data protection, user consent mechanisms, and institutional ethical review as necessary.

\bibliography{custom}

\clearpage
\newpage
\appendix

\section{Survey Instrument}

\begin{tcolorbox}[
  colback=blue!5,
  colframe=blue!100,
  title=Survey Questionnaire (Selected Items),
  boxrule=0.5pt,
  arc=2pt,
  left=6pt, right=6pt, top=6pt, bottom=6pt,
  fonttitle=\bfseries,
  enhanced,
  breakable
]

\textbf{Tool Usage Frequency (Likert scale: Never to Always)}
\begin{itemize}
  \item How often do you use Google for academic tasks?
  \item How often do you use GPT-based tools for academic tasks?
\end{itemize}

\textbf{Perceived Satisfaction, Efficiency, Ease of Use (Likert scale)}
\begin{itemize}
  \item How satisfied are you with the accuracy of results from each tool?
  \item How efficient are these tools in completing academic tasks?
  \item How easy are these tools to use?
\end{itemize}

\textbf{Tool Preference}
\begin{itemize}
  \item Which tool do you prefer overall: Google, GPT, or Both?
\end{itemize}

\textbf{Open-Ended}
\begin{itemize}
  \item In what scenarios do you prefer GPT over Google or vice versa?
  \item What limitations have you faced when using these tools?
\end{itemize}

\end{tcolorbox}

\section{Interview Tasks}

\begin{tcolorbox}[
  colback=blue!5,
  colframe=blue!100,
  title=Assigned Academic Tasks,
  boxrule=0.5pt,
  arc=2pt,
  left=6pt, right=6pt, top=6pt, bottom=6pt,
  fonttitle=\bfseries,
  enhanced,
  breakable
]

Participants were given six structured academic tasks designed to simulate realistic coursework challenges across different disciplines:

\begin{enumerate}
  \item \textbf{Summarize a Research Abstract (All Participants)}\\
  Read a 250-word abstract from a peer-reviewed article and produce a concise 3–5 sentence summary capturing the main objective, methods, and findings.

  \item \textbf{Solve a Coding Problem (CSE Only)}\\
  Write a Python function to compute the factorial of a number, ensuring proper input validation and code documentation.

  \item \textbf{Analyze a Circuit Diagram (EEE Only)}\\
  Interpret a simple resistive circuit with three resistors and a voltage source. Calculate total resistance and current using Ohm’s Law.

  \item \textbf{Interpret a Business Chart (BBA Only)}\\
  Given a bar chart showing quarterly revenue for three products, provide a 5–6 sentence interpretation of trends, anomalies, and business implications.

  \item \textbf{Draft a Formal Email (All Participants)}\\
  Write a professional email to your course instructor requesting an extension on an assignment. The email should be polite, concise, and persuasive.

  \item \textbf{Compare Two Academic Concepts (All Participants)}\\
  Write a short paragraph comparing “quantitative” vs. “qualitative” research methods, highlighting key differences and use cases.
\end{enumerate}

Participants were grouped by tool usage pattern: GPT-only, Google-only, tool-balancing (both sequentially), and random-choice (free selection per task).

\end{tcolorbox}

\section{Task Evaluation Rubric}

\begin{tcolorbox}[
  colback=blue!5,
  colframe=blue!100,
  title=Rubric for Evaluating Task Accuracy (0–10 Scale),
  boxrule=0.5pt,
  arc=2pt,
  left=6pt, right=6pt, top=6pt, bottom=6pt,
  fonttitle=\bfseries,
  enhanced,
  breakable
]

Each academic task was scored on a scale from 0 (poor) to 10 (excellent), based on specific content and skill-based criteria. Rubrics were standardized across evaluators to ensure consistency.

\textbf{1. Summarization Task}
\begin{itemize}
  \item \textbf{Coverage of Key Ideas (0–4):} Accurately identifies main purpose, methods, and findings.
  \item \textbf{Conciseness and Clarity (0–3):} Avoids redundancy; sentences are readable and logically ordered.
  \item \textbf{Language Accuracy (0–3):} Grammar, punctuation, and vocabulary are appropriate for academic tone.
\end{itemize}

\textbf{2. Coding Task (CSE Only)}
\begin{itemize}
  \item \textbf{Correctness (0–4):} Produces correct output for sample inputs.
  \item \textbf{Code Quality (0–3):} Structured, readable, and modular.
  \item \textbf{Input Handling and Comments (0–3):} Includes input validation and descriptive inline comments.
\end{itemize}

\textbf{3. Circuit Analysis Task (EEE Only)}
\begin{itemize}
  \item \textbf{Correct Calculation (0–5):} Accurate application of formulas (e.g., Ohm’s Law).
  \item \textbf{Interpretation and Units (0–3):} Correct labeling and use of units.
  \item \textbf{Clarity of Steps (0–2):} Clear logical progression of calculations.
\end{itemize}

\textbf{4. Business Chart Interpretation (BBA Only)}
\begin{itemize}
  \item \textbf{Insightfulness (0–4):} Accurately identifies trends, anomalies, and patterns.
  \item \textbf{Relevance (0–3):} Comments relate meaningfully to business implications.
  \item \textbf{Clarity (0–3):} Well-structured explanation with clear language.
\end{itemize}

\textbf{5. Formal Email Draft}
\begin{itemize}
  \item \textbf{Professional Tone and Structure (0–4):} Proper salutation, closing, and paragraphing.
  \item \textbf{Persuasiveness (0–3):} Presents a clear and reasonable justification.
  \item \textbf{Grammar and Clarity (0–3):} Language is appropriate, polite, and error-free.
\end{itemize}

\textbf{6. Concept Comparison}
\begin{itemize}
  \item \textbf{Content Accuracy (0–4):} Identifies valid, discipline-appropriate distinctions.
  \item \textbf{Comparative Logic (0–3):} Clearly outlines similarities/differences.
  \item \textbf{Language and Coherence (0–3):} Academic tone and logical flow.
\end{itemize}

\end{tcolorbox}

\section{Thematic Codebook}

Thematic analysis of open-ended survey responses and interview transcripts resulted in four overarching themes. Each theme is described in Table \ref{tab:codebook}, along with its associated codes and representative participant quotes.

\begin{table*}[t]
\centering
\small
\begin{tabular}{p{4cm}p{3.5cm}p{6cm}}
\hline
\textbf{Parent Theme} & \textbf{Child Code} & \textbf{Definition / Quote Example} \\ \hline

\textbf{Effectiveness} 
& Task Fit 
& Tool suitability for academic tasks, which refers to how well a tool matches the academic task at hand. \newline \textit{"GPT is great for summaries, but not so much for detailed citations."} \\ \hline

\multirow{2}{*}{\textbf{Usability \& Cognitive Load}} 
& Ease of Use 
& Simplicity of interaction with the tool, which means how intuitive and straightforward users find the tool. \newline \textit{"ChatGPT saves me time by avoiding extra clicks."} \\

& Information Overload 
& Frustration with excessive irrelevant results, which describes frustration due to excessive, often irrelevant, search results. \newline \textit{"Google gives too many links and I get lost trying to pick one."} \\ \hline

\textbf{Trust and Credibility} 
& Source Verification 
& Need for citable sources, which refers to the extent to which students cross-check the tool output with credible sources. \newline \textit{"I trust Google more when I need something fact-checked."} \\ \hline

\textbf{Contextual Tool Selection} 
& Task Type Influence 
& Decision to use a tool depends on the academic context, which describes tool choice based on the academic context or subject matter. \newline \textit{"For programming help, I use GPT; for research papers, I go with Google."} \\ \hline

\end{tabular}
\caption{Thematic mapping of child codes derived from open-ended survey responses and interviews. Quotes show typical user sentiment for each theme.}
\label{tab:codebook}
\end{table*}

\end{document}